\documentclass[12pt]{article}
\usepackage{a4wide}

\begin{document}
\author{Bernard Linet \thanks{E-mail: linet@lmpt.univ-tours.fr} \\
\small Laboratoire de Math\'ematiques et Physique Th\'eorique \\
\small CNRS/UMR 6083, F\'ed\'eration Denis Poisson \\
\small Universit\'e Fran\c{c}ois Rabelais, 37200 TOURS, France}
\title{\bf Electrostatics in a wormhole geometry}
\date{}
\maketitle
\thispagestyle{empty}

\begin{abstract}
We determine in closed form the electrostatic potential generated by a point
charge at rest in a simple model of static spherically symmetric wormhole. From this,
we deduce the electrostatic self-energy of this point charge.
\end{abstract}

The physical effects in wormholes has been a subject of considerable investigation. The present
work is devoted to solving the electrostatic equation in closed form in the background
metric of Morris and Thorne \cite{morris} which describes a simple wormhole. We take up again the
question  of the electrostatic self-energy analysed by Khusnutdinov and 
Bakhmatov \cite{khus} in the framework of the multipole formalism. We emphasize that we consider
electrostatics in a static spherically symmetric wormhole connecting two different asymptotically 
flat spacetimes. Of course, electromagnetism in a static spherically symmetric
wormhole connecting two distant regions in the same spacetime is radically 
different \cite{frolov}.

In the paper of Morris and Thorne \cite{morris}, a simple geometry of
wormhole is described by the metric
\begin{equation}\label{m1}
ds^2=-dt^2+dl^2+\left( l^2+w^2\right) \left( d\theta^2+\sin^2\theta d\varphi^2 \right)
\end{equation}
in the coordinate system $(t,l,\theta ,\varphi )$ with $-\infty <l<\infty$. The
throat is located at $l=0$. The two different asymptotically flat spacetimes are 
defined for $l\rightarrow \pm \infty$.

To solve in the next the electrostatic equation, it is convenient to write metric (\ref{m1})
in isotropic coordinates $(t,r,\theta ,\varphi )$. The radial cordinate $r$ is
related to the proper radial coordinate $l$ by
\begin{equation}\label{c1}
r=\frac{1}{2} \left( l+\sqrt{l^2+w^2}\right)
\end{equation}
with the range $0<r<\infty$. Now the throat is located at $r=w/2$. We note that
$$
\frac{dr}{dl}=\frac{l+\sqrt{l^2+w^2}}{2\sqrt{l^2+w^2}} \quad
{\rm satisfying} \quad \frac{dr}{dl}>0
$$
and thus coordinate transformation (\ref{c1}) is valid eveywhere. The inverse
transformation is given by
\begin{equation}\label{c2}
l=r-\frac{w^2}{4r} .
\end{equation}
Taking into account (\ref{c1}), metric (\ref{m1}) is written as
\begin{equation}\label{m2}
ds^2=-dt^2+\left( 1+\frac{w^2}{4r^2}\right)^2 \left( dr^2+r^2\left( d\theta^2
+\sin^2\theta d\varphi^2\right) \right)
\end{equation}
in the coordinate system $(t,r,\theta ,\varphi )$ with $r>0$.

We set the Cartesian coordinates $(x^i)$, $i=1,2,3$, associated to the spherical
coordinates $(r,\theta ,\varphi )$. Metric (\ref{m2}) takes then the form
\begin{equation}\label{m3}
ds^2=-dt^2+\left( 1+\frac{w^2}{4r^2}\right)^2 \left( (dx^1)^2+(dx^2)^2+(dx^3)^2\right) 
\end{equation}
with $r=\sqrt{(x^1)^2+(x^2)^2+(x^3)^2}$.

In the background metric (\ref{m3}), we consider a test electromagnetic field generated
by a point charge $e$ at rest. The charge density $\rho$ of this point
charge located at $x_{0}^{i}$ has the expression
\begin{equation}\label{j}
\rho (x^i)=e\frac{\delta^{(3)}\left( x^i-x_{0}^{i}\right)}{\sqrt{-g}}
\end{equation}
where $g$ is the determinant of the metric (\ref{m3}),
$$
\sqrt{-g}=\left( 1+\frac{w^2}{4r^2}\right)^3 .
$$
We note $(r_0,\theta_0,\varphi_0)$ the position of the point charge with $r_0>0$. 

The Maxwell equations in metric (\ref{m3}) for the electric component $A$ of the electromagnetic potential
reduce with source (\ref{j}) to the electrostatic equation
\begin{equation}\label{e}
\triangle A+h(r)\frac{x^i}{r}\partial_iA=-\frac{4\pi e}{1+w^2/4r^2}\,
\delta^{(3)}(x^i-x_{0}^{i})
\end{equation}
where $\triangle$ is the Laplacian operator and the function $h$ has the expression
\begin{equation}\label{h}
h(r)=\frac{1}{1+\displaystyle \frac{w^2}{4r^2}}\frac{d}{dr}\left( 1+\frac{w^2}{4r^2}\right) =
-\frac{w^2}{2r\left( r^2+w^2/4\right) } .
\end{equation}
The aim of this work is to determine in closed form the solution to equation 
(\ref{e}) which is smooth everywhere except of course at the point $x_{0}^{i}$.

In a recent paper \cite{linet}, we have shown that if $h$ obeys certain differential 
relations then there exists a solution to equation (\ref{e}) which can be written in 
the form
\begin{equation}\label{a}
A(x^i)=eNg(r)g(r_0)F(s(x^i,x_{0}^{i}))
\end{equation}
with
\begin{equation}\label{s}
s(x^i,x_{0}^{i})=\frac{\Gamma (x^i,x_{0}^{i})}{k(r)k(r_0)}
\end{equation}
where $\Gamma (x^i,x_{0}^{i})=(x^1-x_{0}^{1})^2+(x^2-x_{0}^{i})^2+(x^3-x_{0}^{3})^2$.
The functions $g$, $k$ and $F$ are to be determined and N is a numerical factor. 

The first step is to determine the elementary function in the Hadamard sense $A_e$
to equation (\ref{e}). This solution is defined by requiring
in the vicinity of the point $x_{0}^{i}$ the following development:
\begin{equation}\label{da}
A_e(x^i)=\frac{1}{\sqrt{\Gamma (x^i,x_{0}^{i})}}\left( U_0(r,r_0)+U_1(r,r_0)
\Gamma (x^i,x_{0}^{i}) +\cdots \right) .
\end{equation}
In the case of metric (\ref{m3}), we have from (\ref{e}) the first coefficient
$$
U_0(r,r_0)=\frac{e}{\sqrt{1+w^2/4r^2}\sqrt{1+w^2/4r_{0}^{2}}} .
$$

By inserting expression (\ref{h}) into the differential relation of our previous
paper \cite{linet}, we obtain
$$
h'+\frac{1}{2}h^2+\frac{2}{r}h= \frac{w^2}{2(r^2+w^2/4)^2} .
$$
In consequence, we know to be able to find a solution in form (\ref{a}) by 
considering the case II of our paper. By comparing, we put
$a=w/2$, $b=w/2$ and $A=-4$. The functions $g$ and $k$ are then given by
\begin{equation}\label{gk}
k(r)=\frac{r^2+w^2/4}{w/2} , \quad g(r)=\frac{r}{r^2+w^2/4},
\end{equation}
and the differential equation for $F$ is
\begin{equation}\label{f}
s(1-s)F''+(-3s+3/2)F'-F=0 .
\end{equation}
According to (\ref{s}) with (\ref{gk}), we have $0\leq s \leq 1$.

We are now in a position to determine in closed form the elementary solution in
the Hadamard sense $A_e$ to equation (\ref{f}) by taking as solution $F_e$
to equation (\ref{f}) 
\begin{equation}\label{fe}
F_e(s)=\frac{1}{\sqrt{s(1-s)}}
\end{equation}
since 
\begin{equation}\label{dfe}
F_e(s)\sim \frac{1}{\sqrt{s}}+\frac{1}{2}\sqrt{s} \quad {\rm as} \quad
s \rightarrow 0 .
\end{equation}
Using (\ref{a}), we get finally
\begin{equation}\label{ae}
A_e(x^i)=\frac{1}{2}ewg(r)g(r_0)
F_e\left( \frac{\Gamma (x^i,x_{0}^{i})}{k(r)k(r_0)}\right) 
\end{equation}
with $N=w/2$ to satisfy (\ref{da}). We can directly verify that $A_e$ given by
expression (\ref{ae}) obeys the partial differential equation (\ref{e}).

However, we are in a case where solution (\ref{ae}) presents another singularity.
Indeed, it defines a new point charge $e$ diametrically opposed to $x_{0}^{i}$ 
on the sphere $r=r_1$ with $r_1=w^2/4r_0$. This corresponds to $s=1$ in expression (\ref{fe}) since
$$
F_e(s)\sim \frac{1}{\sqrt{1-s}} \quad {\rm as} \quad s\rightarrow 1 .
$$

We must look for another solution $F_v$ to equation (\ref{f}) which leads only a
point charge located at $x_{0}^{i}$. We take the solution
\begin{equation}\label{fv}
F_v(s)=\frac{1}{2\sqrt{s(1-s)}}+\frac{\arcsin (1-2s)}{\pi \sqrt{s(1-s)}} .
\end{equation}
The regularity at $s=1$ of expression (\ref{fv}) results from the fact that 
$\arcsin (-1)=-\pi /2$. We note the expansion of function (\ref{fv})
\begin{equation}\label{dfv}
F_v(s)\sim \frac{1}{\sqrt{s}}-\frac{2}{\pi}+\frac{1}{2}\sqrt{s}-\frac{4}{3\pi}s
\quad {\rm as}\quad s\rightarrow 0 .
\end{equation}
This choice of function $F_v$ yields the electrostatic potential
$V$ generated by a point charge $e$ located at $x_{0}^{i}$
\begin{equation}\label{av}
V(x^i)=\frac{1}{2}ewg(r)g(r_0)F_v\left( \frac{\Gamma (x^i,x_{0}^{i})}{k(r)k(r_0)}\right) .
\end{equation}

We point out that the electric flux through the sphere at the infinity
$r\rightarrow \infty$ of the electrostatic potential (\ref{av}) is
\begin{equation}\label{flux}
4\pi e \left[ \frac{1}{2}+\frac{1}{\pi}\arcsin\left( \frac{r_{0}^{2}-w^2/4}
{r_{0}^{2}+w^2/4}\right) \right] .
\end{equation}
We have the opportunity to add to potential (\ref{av}) the electrostatic monopole solution to
equation (\ref{e}) which is smooth everywhere but we have no prescription to fix this
homogeneous solution.

We turn now to calculate the electrostatic self-energy. We develop expression
(\ref{av}) in the vicinity of the point $x_{0}^{i}$ by using expansion (\ref{dfv}). We get
\begin{equation}\label{dav}
V(x^i)\sim \frac{U_0(r,r_0)}{\sqrt{\Gamma (x^i,x_{0}^{i})}} -\frac{ew}{\pi}g(r)g(r_0)
\quad {\rm as}\quad x^i \rightarrow x_{0}^{i} .
\end{equation}
Thus, the electrostatic self-energy $W$ of the point charge $e$ at rest in the
wormhole geometry (\ref{m3}) can be immediately deduces from development (\ref{dav})
\begin{equation}\label{wr}
W(r_0)=-\frac{1}{2\pi}e^2wg^2(r_0) =-\frac{e^2wr_{0}^{2}}
{2\pi \left( r_{0}^{2}+w^2/4\right)^2} .
\end{equation}
The electrostatic self-energy (\ref{wr}) can be expressed with the proper
radial coordinate $l$ in the form
\begin{equation}\label{wl}
W(l_0)=-\frac{e^2w}{2\pi \left( l_{0}^{2}+w^2\right)} 
\end{equation}
Result (\ref{wl}) coincides with the one of Khusnutdinov and Bakhmatov \cite{khus}.

\medskip
I thank Prof. G\'erard Cl\'ement and Nail Khusnutdinov for their friendly correspondence.


\begin{thebibliography}{99}
\bibitem{morris} M. S. Morris and K. S. Thorne {\em Am. J. Phys.} {\bf 56}
395 (1988).
\bibitem{khus} N. R. Khusnutdinov and I. V. Bakhmatov {\em Phys. Rev. D} {\bf 76} 124015 (2007).
\bibitem{frolov} V. P. Frolov and I. D. Novikov {\em Phys. Rev. D} {\bf 42}
1057 (1990).
\bibitem{linet} B. Linet {\em Gen. Rel. Grav.} {\bf 37} 2145 (2005).
\end{thebibliography}
\end{document}